\documentclass[conference]{IEEEtran}

\usepackage{graphicx}
\usepackage[caption=false,font=footnotesize]{subfig}
\usepackage{amsmath}
\usepackage{color}
\usepackage{hyperref}
\usepackage{microtype}
\usepackage{eso-pic}

\AddToShipoutPicture*{\small \sffamily\raisebox{1.8cm}{\hspace{1.8cm}978-1-4577-0351-5/11/\$26.00 \copyright2011 IEEE}}

\begin{document}

\title{Relieving the Wireless Infrastructure: When Opportunistic Networks Meet Guaranteed Delays}

\author{
\IEEEauthorblockN{John Whitbeck$^{1,2}$, Yoann Lopez$^{1}$, J\'er\'emie Leguay$^1$, Vania Conan$^1$, and Marcelo Dias de Amorim$^2$\vspace*{2mm}}
\begin{tabular}{c c c}
$^1$ Thales Communications  & $^2$ CNRS and UPMC Sorbonne Universit{\'{e}}s \\
\end{tabular}}

\maketitle

\begin{abstract}
Major wireless operators are nowadays facing network capacity issues in striving to meet the growing demands of mobile users. At the same time, 3G-enabled devices increasingly benefit from ad hoc radio connectivity (e.g., Wi-Fi). In this context of hybrid connectivity, we propose Push-and-track, a content dissemination framework that harnesses ad hoc communication opportunities to minimize the load on the wireless infrastructure while guaranteeing tight delivery delays. It achieves this through a control loop that collects user-sent acknowledgements to determine if new copies need to be reinjected into the network through the 3G interface. Push-and-Track includes multiple strategies to determine \textit{how many} copies of the content should be injected, \textit{when}, and to \textit{whom}. The short delay-tolerance of common content, such as news or road traffic updates, make them suitable for such a system. Based on a realistic large-scale vehicular dataset from the city of Bologna composed of more than 10,000 vehicles, we demonstrate that Push-and-Track consistently meets its delivery objectives while reducing the use of the 3G network by over 90\%.
\end{abstract}

\IEEEpeerreviewmaketitle

\section{Introduction}
\label{sec:introduction}

In December 2009, mobile data traffic surpassed voice on a global basis, and is expected to continue to double annually for the next five years~\cite{cisco_data,ericsson_data}. Every day, thousands of mobile devices~-- phones, tablets, cars, etc.~-- use the wireless infrastructure to retrieve content from Internet-based sources, creating immense demand on the limited spectrum of infrastructure networks, and therefore leading to deteriorating wireless quality for all subscribers as operators struggle to keep up~\cite{3g_overload}. In order to cool this surging demand, several US and European network operators have either announced or are considering the end of their unlimited 3G data plans~\cite{end_of_all_you_can_eat,3g_orange}.

There are limits however to how much can be achieved by increasing infrastructure capacity or designing better client incentives. Solving the problem of excessive load on infrastructure networks will require paradigm-altering approaches. In particular, when many users are interested in the same content, how can one leverage the multiple ad hoc networking interfaces (e.g., Wi-Fi or Bluetooth) ubiquitous on today's mobile devices in order to assist the infrastructure in disseminating the content? Subscribers may either form a significant subset of all users, comprising for example all those interested in the digital edition of a particular newspaper, or may include all users in a given area, for example vehicles receiving periodic traffic updates in a city.

\begin{figure}
  \centering
  \includegraphics{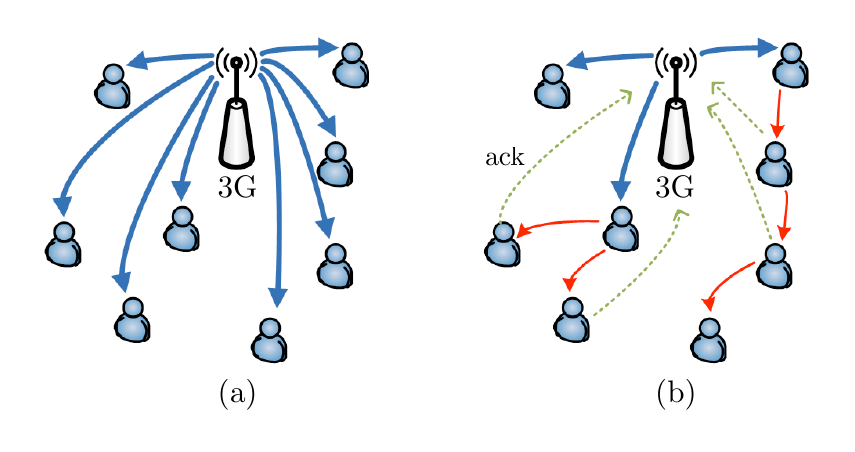}
  \caption{Combining multiple strategies for full data dissemination. Left figure (a) shows the infrastructure-only mode, where the 3G interface is used to send copies of the data to all nodes. In (b), we show the Push-and-Track approach, where opportunistic ad hoc communication is preferred whenever possible. Although acknowledgments are required to keep the loop closed, the global infrastructure load will be significantly reduced.}
  \label{fig:illu}
\end{figure}

In this paper, we address the following question: \textit{how can one relieve the wireless infrastructure using opportunistic networks while guaranteeing 100\% delivery ratio under tight delay constraints?} In particular, we seek to \textit{minimize} the infrastructure load while massively distributing content within a short time to a large number of subscribers. 

We propose \textit{Push-and-Track}, a framework that harnesses both wide-area radios (e.g., 3G or WiMax) and local-area radios (e.g., Bluetooth or Wi-Fi) in order to achieve guaranteed delivery in an opportunistic network while relieving the infrastructure. Our approach is detailed in Fig.~\ref{fig:illu}. A subset of users will receive the content from the infrastructure and start propagating it epidemically; upon receiving the content, nodes send acknowledgments back to the source thus allowing it to keep track of the delivered content and assess the opportunity of \textit{reinjecting} copies. The main feature of Push-and-Track is the closed control loop that supervises the reinjection of copies of the content via the infrastructure whenever it estimates that the ad hoc mode alone will fail to achieve full dissemination within some target delay. To the best of our knowledge, our work is the first to explore this idea.

Unlike accessing an operator's wireless infrastructure, opportunistic forwarding, using short-range ad hoc radio, is essentially free and costs little more than expended battery life. This may not even be a concern in certain circumstances (e.g., vehicular). Unfortunately, it does not provide any guarantees as it depends entirely on the uncontrolled mobility of users.

To this end, we evaluate several reinjection strategies. Push-and-Track splits the problem into \textit{how many} copies of the content should be injected into the network, \textit{when}, and to \textit{whom}. To decide the number of copies to be injected, we define different objective functions of different aggressiveness levels (slow start or fast start). If the dissemination evolution is under the objective, more copies need to be injected through the infrastructure; otherwise, the system remains in ad hoc mode only. For deciding to whom inject copies, we consider randomized, sojourn time, location-based, and connectivity-based strategies.

We thoroughly evaluate all combinations of the proposed strategies by comparing them with both pure infrastructure and pure ad hoc approaches, as well as a near-optimum centralized solution, on a highly realistic large-scale vehicular simulation derived from fine-grained traffic measurements in the city of Bologna. This vehicular dataset is composed of more than 10,000 vehicles covering 20.6~km$^{2}$ and 191~km of roads.

Our results reveal the following findings:

\begin{itemize}

    \item Push-and-Track reduces the infrastructure load by over 90\% when distributing periodic content to all vehicles in the city of Bologna during peak hour traffic while still achieving 100\% on time delivery ratio.

    \item Choosing random recipients for pushing content is a straightforward and efficient strategy.

    \item While always important, reinjection decisions have significantly more impact early in a message's lifetime.

\end{itemize}

\section{Massive dissemination of mobile content with Push-and-Track}
\label{sec:problem}

We consider the problem of distributing dynamic content to a variable set of mobile devices, all equipped with wireless broadband connectivity (3G) and also able to communicate in ad hoc mode. This content is distributed from a point inside the access network infrastructure and can be of any size. Mobile nodes may subscribe to this content based on interest (e.g. news feeds or video podcasts) or for geographical reasons (e.g., road traffic information in my home town). In any case, we assume that the subscriber base is significant enough that ad hoc communication is feasible. We leave the question of users forwarding content they are not interested in open for future work. Furthermore, in this paper, unless specified, we are not concerned with any specific radio technology and will simply refer to \textit{infrastructure} vs. \textit{ad hoc} radios.

Services that are sensitive to jitter, such as VoIP, will of course remain infrastructure-only. Only content that can tolerate some delay in the delivery process (e.g., messages or file transfers) can take advantage of short range communication opportunities. Indeed, they do not have to be downloaded at the instant they are used, and can be smoothly pre-fetched into mobile devices. Most content has an expiration date, either in terms of usefulness for a user (e.g., road traffic information before entering an area), or in terms of validity when updated (e.g., daily news). This expiration date sets the delay-tolerance limit that any dissemination scheme should respect.

Push-and-Track does not rely on any restricted hypothesis on contact statistics. Indeed, many opportunistic routing schemes require a learning or bootstrapping phase during which nodes aggregate statistics about meeting probabilities~\cite{lindgren03}. In particular, a lot of attention has been focused on pairwise contact and inter-contact time distributions. These may be relevant in certain very specific circumstances, such as a conference, in which people regularly meet and separate, but are much less relevant in an urban vehicular context for example, where nodes typically meet only once. Furthermore, in a real system, users expect to be able to access the content immediately, not after some learning period. Any general realistic opportunistic content dissemination scheme which aims at guaranteeing delays cannot therefore rely only on statistical knowledge of node mobility and behavior.

Push-and-Track is a mobility-agnostic framework for massively disseminating content to mobiles nodes while meeting guaranteed delays and minimizing the load on the wireless infrastructure. It consists of a control system which \textit{pushes} periodical content to mobiles nodes and \textit{keeps track} of its opportunistic dissemination. It uses a closed-loop controller to decide at each time step $\Delta_t$ which nodes should receive the content from the infrastructure (push operation) to ensure a smooth and effective dissemination using epidemic routing. Upon receiving the content, each node sends an acknowledgement back to the control system using the infrastructure network. This allows the controller to keep track of the remaining nodes to serve. By designing the system in a way that this feedback information is much smaller than the content itself, we expect to obtain significant reduction of the traffic flowing through the 3G infrastructure.

\section{Reducing infrastructure load: strategies}
\label{sec:strategies}

The content is propagating among the mobile subscribers, acknowledgments are coming in, the deadline is approaching: should copies be reinjected into the network? If so, how many and to whom? Guaranteeing 100\% delivery ratio while minimizing the load on the infrastructure is the heart of Push-and-Track. Each reinjection strategy therefore consists of two parts. At every time step, it will first determine how many, if any, copies must be reinjected, and then determine for each new copy whom to push it to.

\subsection{Assumptions}
\label{subsec:assumptions}

A content is issued at time $t_i$ and must be delivered to all target nodes within a period of $T$ seconds. Nodes may enter in the system in the middle of a period but they should receive the message before its expiration. Push-and-Track slots period $T$ into time steps of $\Delta_t$ seconds that correspond to the instants the feedback loop controlling the dissemination process decides whether or not to reinject new copies of the content. The dissemination process operates by pushing content to a subset of non infected nodes.

\subsection{Reference strategies}
\label{subsec:benchmarking}

The strategies developed in this section will be compared to the following upper and lower limits on achievable performance:

\smallskip\noindent\textbf{Infrastructure only:} All content is pushed exclusively through the infrastructure. No ad hoc communications are allowed. This represents the baseline cost of massive content distribution using present-day deployments.

\smallskip\noindent\textbf{Dominating set oracle:} All content is pushed to a small number of precalculated nodes. For each message, we define a directed graph, in which each vertex is connected to all the vertices to which there exists a space-time forwarding path during the message's lifetime. The infrastructure then pushes the content to a dominating set for this graph.\footnote{Here, a \textit{dominating set} is a set of nodes in the directed graph such that each node is either in the dominating set or has an inbound edge from a node in the dominating set.}  This is analogous to the well known problem of choosing multipoint relays for broadcasting in a wireless network~\cite{laouiti2001}. Finding a minimal dominating set is NP-complete but a simple greedy algorithm provides a dominating set whose cardinality is at most $\log K$ times larger than the optimal set, where $K$ is the maximum degree of a node in the aforementioned graph~\cite{laouiti2001}. Results obtained by pushing content exclusively to nodes in this dominating set constitute our performance target.

\subsection{When to push}
\label{subsec:when}

\begin{figure}
  \centering
  \includegraphics{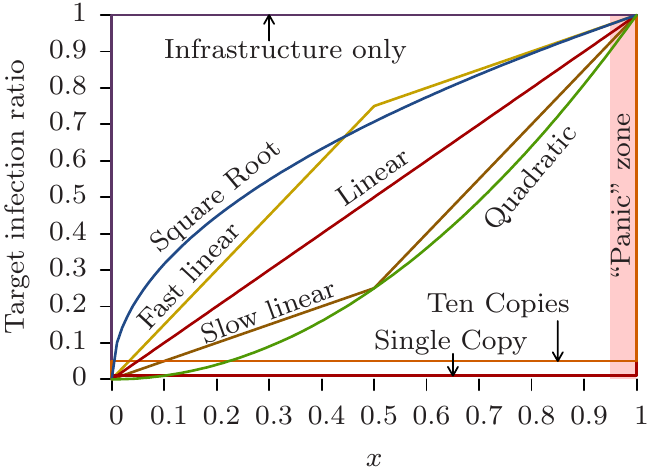}
  \caption{Infection rate objective functions. $x$ is the fraction of time elapsed between a message's creation and expiration dates. $x=1$ is the deadline for achieving 100\% infection.}
  \label{fig:objective_functions}
\end{figure}

Is it better to inject a small number of initial copies, and run the risk of having to push large numbers of copies as the deadline approaches, or jump-start the epidemic dissemination with many initial copies, despite the fact that some of those may turn out to be redundant? How about keeping a steady reinjection rate over the course of a message's lifetime? The strategies outlined in this section, hereafter called \textit{when-strategies}, cover all these questions.

Let $x$ be the fraction of time elapsed between a message's creation and expiration dates. Each strategy is defined by an \textit{objective function} (see Fig.~\ref{fig:objective_functions}), which indicates for every $0 \le x \le 1 $ what the current \textit{infection ratio} should be (i.e., the fraction of the number of subscribing nodes that have the content). Note that the infection ratio can go down if nodes unsubscribe. If, at any time, the measured infection ratio, obtained from the acknowledgments, is below the current target infection ratio, then the strategy returns the minimal number of additional copies that need to be reinjected in order to meet that target. Furthermore, when the time left before the deadline is equal to the time required to push the message directly through the infrastructure, the control system enters a ``panic zone'' (Fig.~\ref{fig:objective_functions}) in which the infrastructure pushes the content to all nodes that have not yet received it.

The \textit{when-strategies} may broadly be divided into three categories:

\smallskip\noindent\textbf{Slow start:} This includes two very simple ``push-and-wait'' (in opposition to Push-and-Track) strategies that push an initial number of copies and then do nothing until the panic zone: the \textit{Single Copy} and \textit{Ten Copies} strategies which respectively inject one and ten initial copies. The objective function for the \textit{Quadratic}, or ``very slow start'', strategy is $x^2$. The \textit{Slow Linear} strategy starts with a $\frac{x}{2}$ linear objective for the first half of the message's lifetime, and finishes with a $\frac{3}{2}x-\frac{1}{2}$ objective.

\smallskip\noindent\textbf{Fast start:} The objective function for the \textit{Square Root}, or ``very fast start'', strategy is $\sqrt{x}$. The \textit{Fast Linear} strategy starts with a $\frac{3}{2}x$ linear objective for the first half of the message's lifetime, and finishes with a $\frac{x}{2}+\frac{1}{2}$ objective.

\smallskip\noindent\textbf{Steady:} This is the \textit{Linear} strategy which ensures an infection ration strictly proportional to $x$.

\subsection{To whom}
\label{subsec:whom}

Once the number of copies to reinject has been decided, the next question is whom to push it to. In this paper we test the following \textit{whom-strategies}:

\smallskip\noindent\textbf{Random:} Push to a random node chosen uniformly among those that have not yet acknowledged reception.

\smallskip\noindent\textbf{Entry time:} If content subscription is localization-based, then each node's entry time (i.e., subscription time) is correlated to its position in the area. For example, pushing to those that have the most recent (\textit{Entry-Newest}) or oldest (\textit{Entry-Oldest}) entry times should target nodes close to the edge of the area, whereas pushing to those that are closest to the average entry time (\textit{Entry-Average}) should target the middle of the area.

\smallskip\noindent\textbf{GPS-based:} On top of the existing control messages, each node may also periodically inform the control system of its current location. From this information, the space encompassing all nodes is recursively partitioned according to the Barnes-Hut method~\cite{barneshut86}. The idea is to keep on dividing each rectangular area into four sub-areas until either an area has only one node in it, or a maximum recursion level has been reached. This allows efficient computations of node density and force-based algorithms. In this paper, two GPS-based strategies were considered. In order to ensure rapid replication, \textit{GPS-Density} pushes the content to an uninfected node within the highest density area. In \textit{GPS-Potential}, each infected node $i$ applies to every other node $j$ a Coulomb potential equal to $\frac{1}{d_{ij}}$ ($d_{ij}$ is the distance between $i$ and $j$). Each side of the space also creates a potential equal to that of a single infected node. In order to spread the copies as well as possible over the entire space, \textit{GPS-Potential} pushes the content to the node with the lowest potential.

\smallskip\noindent\textbf{Connectivity-based:} Ad hoc routing protocols try to provide each node with a good enough picture of the global network topology to make intelligent routing decisions. On the other hand, opportunistic routing protocols only assume knowledge of the current neighbors. However, nodes can periodically communicate to the control system a list of their current neighbors. Even though each node will still perform opportunistic store-and-forwarding, the control system will have a good slightly out of sync, picture of the global connectivity graph. The \textit{CC} (Connected Components) strategy uses this information to push content to a randomly chosen node within the largest uninfected connected component. If all connected components have at least one infected node, then it pushes to a node within the one with the most uninfected nodes. The idea is to push only one copy per connected component thereby getting close to the optimal number of pushed copies.

\subsection{Control loop operation}
\label{subsec:control-loop}

The control loop is the core of the decision system. The infrastructure must be aware of the dissemination status at all times to decide or not to inject new copies of the data in the network. To this end, the following control messages are mandatory. In the vehicular scenario described in the next Section, each vehicle sends an \textsf{ENTER} message (i.e., subscribe) upon entering the simulation area and a \textsf{LEAVE} message (i.e., unsubscribe) upon leaving. As soon as a vehicle receives the data, it sends an \textsf{ACK} message (i.e., acknowledgment) back to the control system.

\section{Bologna vehicular dataset}
\label{sec:dataset}

Many existing datasets were considered for evaluating Push-and-Track, in particular the Bluetooth contact traces obtained in a conference~\cite{chaintreau}, on a campus~\cite{mit}, or during a rollerblading tour~\cite{tournoux08}. Unfortunately, these all have a small fixed set of participants (roughly 100) and the underlying social affinities and dynamics translate into specific inter-contact patterns that have a crucial impact on data dissemination. For our purposes, we wanted a realistic dataset with a large variable number of users and a high turnover rate among the users to simulate subscription and unsubscription. Furthermore, as in real-life, we expect those users to be mostly strangers to each other, and therefore wished to keep social dynamics to a minimum. The Bologna vehicular dataset described in this section has all these features.

\subsection{Dataset construction}

\begin{figure}
    \centering
    \includegraphics[scale=0.5]{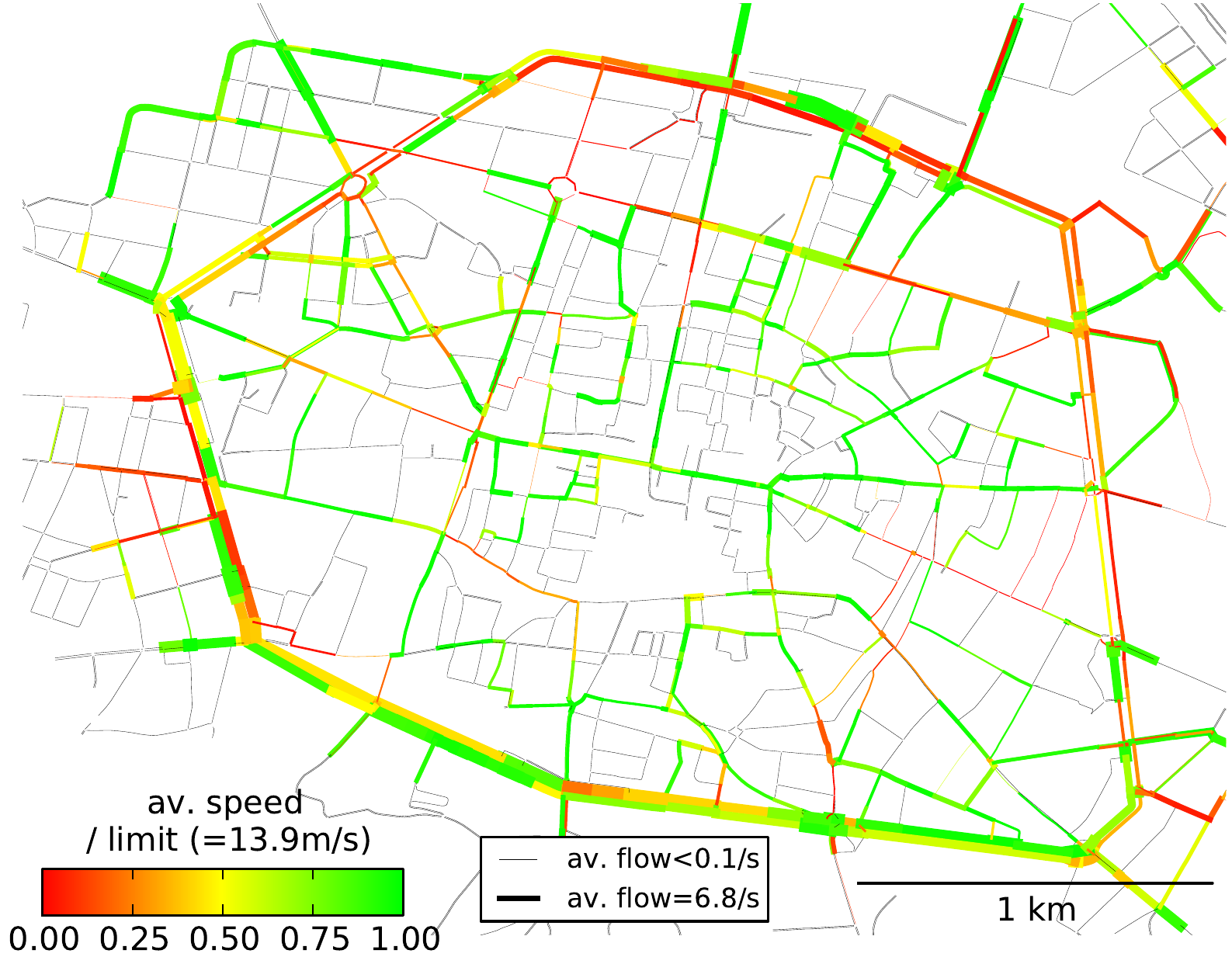}
    \caption{Simulated Bologna road network. The incoming vehicle flow on a given lane is anamorphically represented by its width on a logarithmic scale. The vehicle speed on a lane is represented by a color gradient. Both values are averaged over the duration of the dataset.}
    \label{fig_map}
\end{figure}

We evaluate Push-and-Track on a large-scale vehicular mobility excerpt of a city-wide dataset of the municipality of Bologna (Italy). This dataset's initial purpose was to evaluate future cooperative road traffic management strategies within the iTetris European project~\cite{itetris}. In this paper, we focused on the area surrounding Bologna's city center, displayed in Fig.~\ref{fig_map}, covering $20.6$ km$^2$ and including $191$ km of roads. It exhibits diversity in terms of road types: a ring-shaped main road yields to various entry points to the historical city center.

The dataset is derived from measurements of traffic conditions realized by the municipality of Bologna on their road network. The supplied raw data includes measurements of circulating vehicles acquired by 636 induction loops spread over the city and a synthesis of user surveys on usual commuting trips. Exploiting this raw data, the OD (Origin-Destination) vehicle flow matrices yield macroscopic traffic demands on the city road network during common weekday peak hours (from 8~a.m. to 9~a.m.). Monday and Friday mornings were discarded to avoid specific traffic patterns due to week-end proximity.

Using common traffic engineering tools~\cite{vissim}, the macroscopic traffic demands and route assignments are then used to infer individual vehicle micro-mobility on a highly-accurate representation of the Bologna road network. We ran the simulation with SUMO, an open-source microscopic vehicular movement simulator generally used by the vehicular research community for testing and comparing models of vehicle behavior, traffic light optimization, and vehicle routing~\cite{krajzewicz2002sumo}. To model individual vehicle behavior, SUMO uses a space-continuous and time-discrete car-following model on a multiple-lane road network representation~\cite{zpr98-319}. The latter is supplied in the Bologna dataset and includes traffic lights' positions and patterns, lane-changing, and junction-based right-of-way rules. Macro and micro mobility videos are available online~\cite{pnt}.

\subsection{Dataset analysis}

\begin{figure}
  \centering
  \subfloat[CCDFs for contact and transit times.\label{subfig:dataset_cdfs}]{
    \includegraphics{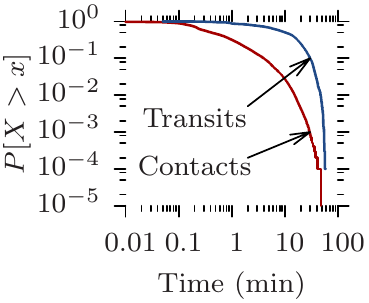}} \quad
  \subfloat[Number of connected components (CC).\label{subfig:numcc}]{
    \includegraphics{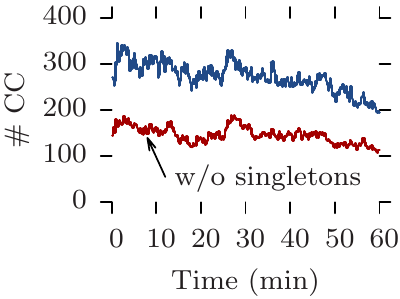}}
  \caption{Characteristics for the Bologna Ringway dataset.}
  \label{fig:dataset_analysis}
\end{figure}

We now analyze the vehicular traffic and network connectivity statistics of the simulation. After a warm-up period, the traffic is simulated during 3,600~s, which leads to a total number of 10,333 simulated vehicles. During this hour, a maximum of 4,494 and an average of 3,540 vehicles are simultaneously present on the road network. As in real-life, traffic conditions vary from fluid to congested in different parts of the city. This is reflected in the vehicles' transit times (Fig.~\ref{subfig:dataset_cdfs}). Indeed, vehicles remain an average of 13.2 minutes in the city area. While most of these are short trips (50\% are below 10 minutes), some last for over 50 minutes long. Fig.~\ref{fig_map} shows the number of vehicles and average speeds on each road in Bologna. It highlights the relatively larger amount of traffic on the surrounding ring-shaped multiple-lane road than on the capillary network, which is mainly single-lane. Due to dense morning traffic, right-of-way rules, and traffic lights, traffic jams occur on the outer ring and at crossroads.

We define a contact as a robust communication that allows reliable data delivery between two vehicles. We assume that all the vehicles may communicate in an ad hoc fashion using the IEEE~802.11 amendment for Wireless Access in Vehicular Environments (WAVE)~\cite{80211p}. As wireless propagation models are not the core of this paper, we assume a deterministic model where a packet is successfully received if the receiver's distance is below a certain indicative value. Following a pragmatic approach, we consider path loss model approximations and measurements in a urban line-of-sight environment performed by Cheng et al.~\cite{cheng2007mobile}, both corroborating on the existence of a critical distance at $d=100$~m, above which radio propagation suffers from high degradation and variability. Vehicles less than $100$~m apart were therefore considered within transmission range of each other. The resulting network contact duration distribution is illustrated in Fig.~\ref{subfig:dataset_cdfs}. Up to 5 minutes, the distribution may be approximated by a power-law before following an exponential decay. Most contacts are short lived (50\% last less than 25 seconds), illustrating the highly dynamic nature of the vehicular mobility, but a few last up to 50 minutes.

We define the connectivity graph as a time-variant undirected graph with mobile nodes as vertices. Mobile nodes are connected if a contact exists between them. The evolution of the number of connected components in the connectivity graph is depicted in Fig.~\ref{subfig:numcc}. Despite the important number of vehicles and the presence of some large connected components (up to 1,200 nodes), the network remains highly partitioned at all times with a large amount of isolated vehicles. In good opportunistic fashion, exploiting node mobility is therefore crucial to achieving connectivity over time.

\section{Simulation results}
\label{sec:results}

\subsection{Simulator}

The results in this section are all based on the Bologna car traffic dataset from a typical weekday between 8~a.m. and 9~a.m. described in the previous section. Unfortunately, none of the existing network simulators we surveyed~\cite{ONE,ns3}, on top of severe scalability issues when simulating several thousand users, were adapted to evaluate Push-and-Track strategies. For the purposes of this paper, we built our own simulator, heavily inspired by the ONE DTN simulator~\cite{ONE}. In particular, it retains the contact-based ad hoc communication model from ONE, with its simple interference model in which a node may only communicate with a single neighbor at the same time. Unlike ONE, all routing is broadcast, there are different classes of messages (e.g., content or control), and different wireless media (e.g., infrastructure and ad hoc). Furthermore, we assume that each user has a non-interfering infrastructure link to the control system with different upload and download rates.

Vehicles send \textsf{ENTER}, \textsf{LEAVE}, and \textsf{ACK} control messages as described in Section~\ref{subsec:control-loop}. As for the optional messages, we set a timer of one minute for both the GPS-based and Connected Components strategies.

All transfers, including control messages, are simulated and may fail. An ad hoc transfer will fail if either the nodes move out of range of each other or one of the nodes leaves the area before the end of the transfer. An infrastructure transfer, with the exception of the \textsf{LEAVE} messages, will also fail if the node leaves the area too early. Furthermore, a node may be simultaneously receiving the same message from both the infrastructure and directly from another node; whichever one finishes first cancels the other. The amount transferred before the cancel of course counts against the total loads for ad hoc or infrastructure.

\subsection{Experimental setup}

As in any simulation, there are a number of parameters whose values inevitably incur some arbitrariness. We tried to keep this to a minimum. The bit-rate of the ad hoc links is set to 1~Mbytes/s which is compatible with the IEEE~802.11 amendment for wireless in vehicular environments (WAVE)~\cite{80211p}. The bit-rate for the infrastructure downlink is set to 100~Kbytes/s. This is double the expected bit-rate of EDGE networks but much less than the advertised 7.2~Mbits/s rate of HSDPA. However, surveys in Europe and the US have shown that the average user-experienced 3G downlink rate is typically just below 128~Kbytes/s~\cite{wired_3g,ufc_3g}. The infrastructure uplink rate is set to 10~Kbytes/s. Furthermore, each content message is set to 1~Mbyte in size. This means that it takes 10 seconds to transfer over the infrastructure and 1 second over the ad hoc link. The bit-rates that we consider here might either be optimistic or pessimistic depending on nodes location, velocity, or on the access networks they use. Because our evaluation is meant to demonstrate how Push-and-Track can leverage opportunistic communications, we make simplistic assumptions on low layers, and leave more accurate evaluations for future work. Finally, for the sake of simplicity, control messages are all 256-bytes long. This is probably excessive for simple \textsf{ENTER}, \textsf{LEAVE}, and \textsf{ACK} messages, but long enough to accommodate a sizable list of neighbors. The load induced by control messages is of course included in the total infrastructure load but is typically one or more orders of magnitude less than the load incurred by pushing the content to nodes. The control loop's time step $\Delta_t$ was set to 0.01 seconds.

Even though our simulator can handle multiple competing messages, in order to properly identify the important factors influencing message propagation, we limited ourselves to a single message at any given time in the network. In practice, messages are sent periodically, with the previous one expiring as the next one is sent. In this paper, two message lifetime periods were tested: a tight 1-minute delay and a more relaxed 10-minute delay. As we will see, the results differ significantly between these two constraints.

Each pair of \textit{when} and \textit{whom} strategies, described in Section~\ref{sec:strategies}, were tested. A run spans the full hour of the dataset and consists in periodically sending a new message and then controlling its propagation using a particular strategy pair. In this paper, due to space constraints, we only present a small subset of our results. This section presents two types of results: global averages (Figs.~\ref{fig:infra_relief}, \ref{fig:beating_random}, and~\ref{fig:vs_freeze}) and dynamic averages (Figs.~\ref{fig:dyn60} and~\ref{fig:dyn60_freeze}). The global results are averages over 10 runs. In order to smooth out effects due to the particular network topology at the beginning of each period, the sending time of the first message is shifted by $T/10$ at every subsequent run, where $T$ is the sending period (i.e., the message lifetime). The dynamic results are also averages over 10 runs but are focused on a specific period and hence without any shifting of the sending time of the first message. 95\% confidence intervals were calculated for every measurement. These are typically very tight. A video of the Quadratic Random strategy is available online~\cite{pnt}.

\subsection{Relieving the infrastructure}
\label{subsec:relief}

\begin{figure}
  \centering
  \includegraphics{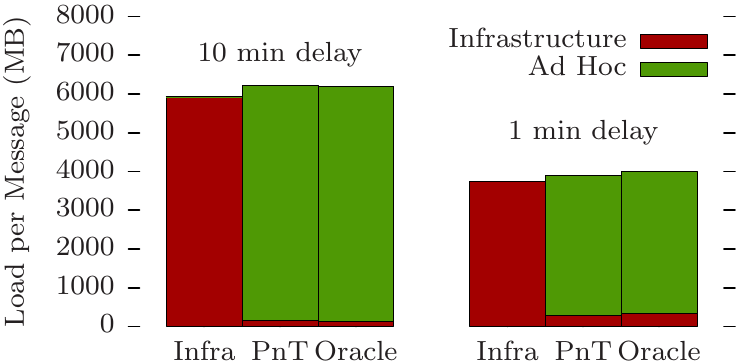}
  \caption{Infrastructure vs. ad hoc load per message sent using only the infrastructure (Infra), Push-and-Track (PnT), and the Dominating Set Oracle (Oracle).}
  \label{fig:infra_relief}
\end{figure}

Push-and-Track does an excellent job of relieving the load on the infrastructure by transferring most of it to faster and cheaper ad hoc communications. Fig.~\ref{fig:infra_relief} shows the average total amount of information transferred per message and how this is split between infrastructure and ad hoc. The results for Push-and-Track correspond to the best \textit{when} and \textit{whom} pair of strategies for a 10-minute delay (\textit{Slow Linear~/ CC}) and a 1-minute delay (\textit{Quadratic~/ CC}). The following sections will examine how the different strategies combine in more detail. The totals for a 10-minute delay are greater than those for a 1-minute delay. Recall that most vehicle transit times are less than 10 minutes (see Section~\ref{sec:dataset}). Therefore, there are more vehicles in the simulation area over a 10-minute period than a 1-minute period, hence the difference in total transfer amounts per message.

Push-and-Track manages to transfer nearly all of the load from the infrastructure to ad hoc communications: 97\% for a 10-minute delay, and 92\% for a 1-minute delay. The ratio is less good with a tighter delay simply because the epidemic ad hoc dissemination has less time to propagate the message to the entire network and thus more copies must be reinjected to parts of the network that have not yet received the content.

Furthermore, with a 10-minute delay, Push-and-Track only exceeds by 28\% the infrastructure load obtained through the \textit{Dominating Set Oracle}. With a long delay, the epidemic propagation has time to fully explore every space-time path. Therefore pushing a small number of initial copies to a good dominating set of the spatial-time directed graph is a very difficult strategy to beat.

Interestingly, with a tighter 1-minute constraint, Push-and-Track actually outperforms the \textit{Dominating Set Oracle} by 13.5\%. There are several reasons for this. Firstly, recall that the dominating set is that of a special directed graph in which each vertex is connected to all the vertices to which there exists a space-time forwarding path during a message's lifetime. The dominating set calculated by the oracle is not a minimum dominating set of this graph, but its cardinality is within a $\log K$ factor of that of the minimum dominating set, where $K$ is the maximum degree of a node in the graph (see Section~\ref{subsec:benchmarking}). However $K$ can be quite large in our experiment (up to roughly 1,500), thus $\log K \approx 7$. Put differently, the minimum dominating set could be up to 7 times smaller than the one calculated by the oracle.

Secondly, and much more importantly, the epidemic propagation does not have time to fully explore every space-time path within 1 minute. For example, if a node from a large connected components moves to another large connected component late during the 1-minute period, the oracle will assume there exists a space-time path from any node in the first connected component to any node in the second one. However that does not mean that by injecting one copy into the first connected component, that everyone in the second connected component will be infected before the end of the message's lifetime. This means that the oracle hits the ``panic zone'' (see Section~\ref{sec:strategies}) before having infected every node. Whatever efficiency is gained by an excellent choice of initial nodes to infect is lost when it has to push the content to all remaining uninfected nodes as the deadline gets close. On the other hand, Push-and-Track, by keeping track of the epidemic's progression and reinjecting copies when needed, is less affected by the ``panic zone'' and thus can outperform the oracle despite making poorer choices of whom to push to. This underscores the main point of this paper: \textit{having a feedback loop for reinjecting content is essential for guaranteeing delivery delays in a hybrid infrastructure/ad hoc network}.

\subsection{Beating random}
\label{subsec:beating_random}

When surveying the results for all \textit{when} and \textit{whom} strategy pairs, the \textit{Random} reinjection strategy consistently does better than most of the more sophisticated strategies described in Section~\ref{sec:strategies}. This section examines this observation in more detail and studies the impact of \textit{whom}-strategies on the infrastructure load.

\begin{figure}
  \centering
  \includegraphics{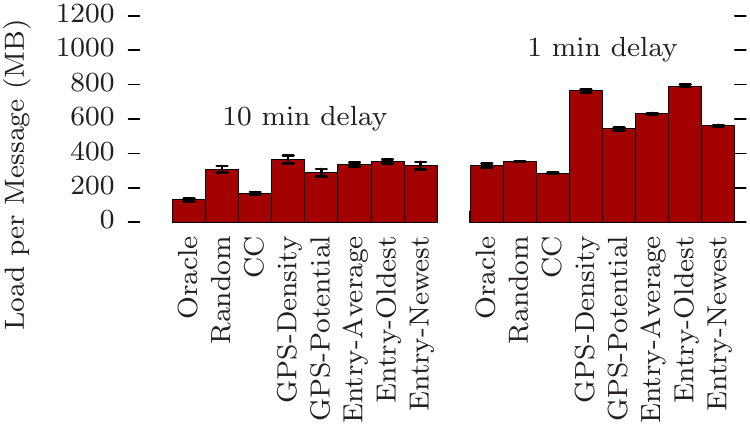}
  \caption{Infrastructure load per message for different \textit{whom}-strategies. Each set of results uses its best \textit{when}-strategy for reinjection: \textit{Slow linear} for the 10 min results and \textit{Quadratic} for the 1 min results. 95\% confidence intervals are shown on top of each bar.}
  \label{fig:beating_random}
\end{figure}

Fig.~\ref{fig:beating_random} plots the average infrastructure load per message for different \textit{whom}-strategies. Each set of results uses its best \textit{when}-strategy for reinjection: \textit{Slow linear} for the 10-minute results and \textit{Quadratic} for the 1-minute results. The load measurements include the control load. With a 10-minute delay, this amounts to roughly 3~Mbytes per message, except for the \textit{GPS-based} and Connected Components (\textit{CC}) strategies, where it goes up to 15~Mbytes per message due to the periodic updates on current position or current neighbors. With a 1-minute delay, those numbers become 1~Mbyte and 2~Mbytes, respectively. In any case, they remain small compared to the load on the downlink.

The results for the 10-minute delay on Fig.~\ref{fig:beating_random} reinforce the previous section's observation that given enough time pushing even a single copy to any node in the area will be sufficient. In this case, the only strategy that significantly outperforms \textit{Random} is the \textit{CC} strategy. Here, the extra overhead incurred by the extra control messages is clearly worth the effort.

With a 1-minute delay, nearly every strategy performs significantly worse than \textit{Random}. In particular, the \textit{GPS-Density} strategy frequently targets nodes that are both in the same dense connected component, leading to many ``useless'' pushes. The \textit{GPS-Potential} improves on this by spreading the copies to the least infected areas, but, because of this, will frequently push to nodes in areas of very sparse connectivity. The \textit{Entry-Newest} and \textit{Entry-Oldest} tend to target nodes on the edge of the simulation area, whereas the \textit{Entry-Average} targets node closer to the center. \textit{Random} combines the best of all these strategies. Indeed it statistically has a high chance of hitting the large connected components and also tends to spread the copies uniformly over the area. Again the only strategy that beats it is the \textit{CC} strategy.

If one is not willing to deal with the added complexity of a more sophisticated control channel, let alone privacy concerns about localization and/or proximity information, then the simple \textit{Random} \textit{whom}-strategy consistently performs very well.

\subsection{Fast or slow start?}
\label{subsec:fast_slow}

We examine how the infection ratio evolves over the course of one message's lifetime for different \textit{when-strategies}. All results in this section use the \textit{Random} \textit{whom-strategy}. What is the better strategy: sending many initial copies, in order to avoid the ``panic zone'', or few, at the risk of having to push extra copies as the deadline get close?

\begin{figure}[t]
  \centering
  \subfloat[Fast start: (1) Square root, (2) Fast linear, (3) Linear.\label{subfig:dyn60_fast}]{\includegraphics{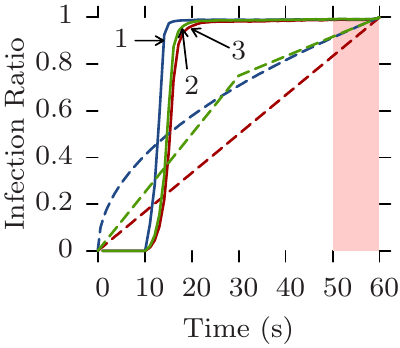}} \quad
  \subfloat[Slow start: (3) Linear, (4) Slow linear, (5) Quadratic, (6) Ten copies, (7) Single copy.\label{subfig:dyn60_slow}]{\includegraphics{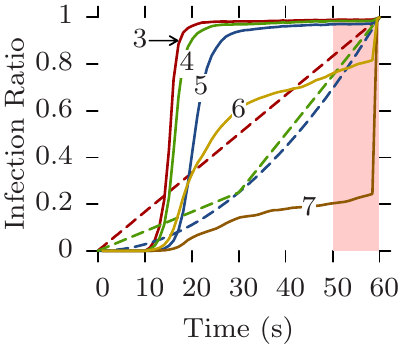}}
  \caption{Infection rates with 1-minute maximum delay depending on the when-strategy. All results are for the \textit{Random} reinjection strategy. Objective functions are dashed and the light red area corresponds to the ``panic zone''.}
  \label{fig:dyn60}
\end{figure}

Fig.~\ref{fig:dyn60} shows the evolution of the infection ratio for various slow-start and fast-start strategies with a 1-minute delay. The corresponding objective functions are represented by dashed lines and the panic zone is the light red area. On both Figs.~\ref{subfig:dyn60_fast} and~\ref{subfig:dyn60_slow}, the infection ratio is zero for the first ten seconds, which is the time required to send a copy over the infrastructure. However, from the point of view of the control system, a node is considered infected as soon as a transfer is initiated to avoid any explosion in the number of initiated transfers. Therefore during the initial ten seconds, from the point of view of the control system, the infection ratio is exactly equal to the target ratio. Once the epidemic propagation kicks in, the real infection ratio grows rapidly. For the quick-start strategies in Fig.~\ref{subfig:dyn60_fast}, this means achieving an infection ratio of nearly 1 after only 20 seconds. For the slow-start strategies in Fig.~\ref{subfig:dyn60_slow}, the \textit{Slow linear} strategy is in fact nearly as fast as the \textit{Linear} strategy. The \textit{Quadratic} strategy slows down the infection ratio and achieves near-complete coverage after about 40 seconds. On the other hand, the \textit{Ten Copies} and the \textit{Single Copy} strategies fail to achieve complete coverage before entering the ``panic zone'' and therefore must reinject many copies at the end.

The latency of the infrastructure links (10 seconds in our example) imposes a delay between the moment when a reinjection decision is taken, and the moment when that decision has an effect on the epidemic propagation. This is particularly tricky during the first 10 seconds when no copies have yet begun disseminating in the ad-hoc network. During that time, the feedback loop is essentially blind. The steep slopes of Fig.~\ref{fig:dyn60} suggest that, even for the slow-start strategies, Push-and-Track may be overreacting during those initial seconds.

\begin{figure}[t]
  \centering
  \subfloat[Fast start: (1) Square root, (2) Fast linear, (3) Linear.\label{subfig:dyn60_freeze_fast}]{\includegraphics{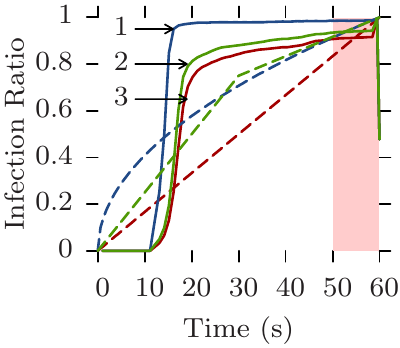}} \quad
  \subfloat[Slow start: (3) Linear, (4) Slow linear, (5) Quadratic, (6) Ten copies, (7) Single copy.\label{subfig:dyn60_freeze_slow}]{\includegraphics{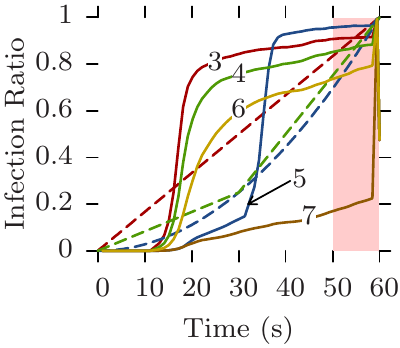}}
  \caption{Including a \textit{freezing} mechanism in the feedback loop: infection rates with 1-minute maximum delay depending on the when-strategy. All results are for the \textit{Random} reinjection strategy. Objective functions are dashed and the light red area corresponds to the ``panic zone''.}
  \label{fig:dyn60_freeze}
\end{figure}

In order to test this hypothesis, we modify the feedback loop with a \textit{freezing} mechanism. While a message is ``frozen'', the infrastructure will not push it to anyone. Each time the infrastructure pushes a batch of new copies, the message is frozen for a period equal to twice its transmission time (20 seconds in our example). This guarantees that the infrastructure doesn't trigger a new reinjection until the previous one has had time to make an impact. Furthermore, to prevent every strategy, fast-start or slow-start, from freezing the messages after sending a single initial copy right at the very beginning of a period, each new message is initially frozen for 1 second. For example, after that 1 second, a \textit{Square Root} strategy will inject more initial copies than a \textit{Quadratic} one.

Fig.~\ref{fig:dyn60_freeze} plots the same dynamic infection ratios as Fig.~\ref{fig:dyn60} but with the freezing mechanism. As expected, the infection rates for all strategies have been slowed down, while still allowing the system enough time to react. The case of the \textit{Quadratic} strategy is very interesting. Because it starts so slowly, it initially sends a single copy before freezing. The epidemic propagation started by that single copy is not fast enough to catch up with the objective function by the end of the freezing period. It then overreacts by sending too many copies to catch up and its infection ratio then overtakes that of supposedly faster strategies.

\begin{figure}
  \centering
  \includegraphics{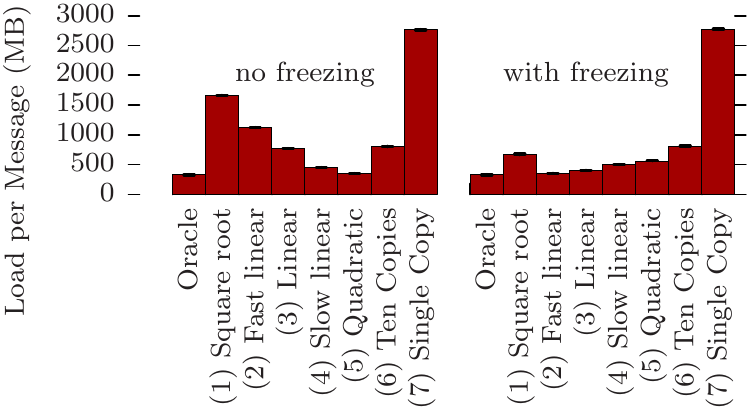}
  \caption{Infrastructure load per message for different when-strategies. All results are for the \textit{Random} reinjection strategy with a 1-minute maximum delay. 95\% confidence intervals are shown on top of each bar.}
  \label{fig:vs_freeze}
\end{figure}

Intuitively, we expect the freezing strategies to send fewer copies on the infrastructure than their non-freezing counterparts. This is broadly true but with a little twist. Fig.~\ref{fig:vs_freeze} compares the total infrastructure load per message for the freezing and the non-freezing strategies. Interestingly, slow-start strategies perform better with no freezing, but, with the exception of the \textit{Square Root} strategy, the reverse is true when using the freezing mechanism. The best strategy is \textit{Quadratic} without freezing but \textit{Fast Linear} with it. A close look at Figs.~\ref{subfig:dyn60_slow} and~\ref{subfig:dyn60_freeze_fast} shows that the infection rates for these two strategies nearly overlap.

We make several interesting observations out of these results. It seems that the \textit{crucial reinjection decisions occur the very beginning of a message's lifetime}. The earlier a copy is sent, the more time it has to have an impact. Copies sent during the epidemic phase-transition are nearly useless. Later, it seems preferable to just wait for the panic zone rather than proactively adding new copies. The goal therefore is ensure that enough copies are present early on to trigger the epidemic phase-transition but not to overdo it and uselessly burden the infrastructure.

\section{Related work}
\label{sec:related}

Reducing the load on the wireless infrastructure has received attention in both academic and industrial circles. For example, Balasubramanian et al. exploit the delay-tolerance of common types of data such as emails or file transfers to opportunistically offload them to available open Wi-Fi hotspots~\cite{balasubramaniam:augmenting}. The now defunct French MVNO Ten Mobile had been offering free pushes of podcasts to their customers' mobile phone during the night using cheaper minutes~\cite{ten_mobile}. Every morning, users had the latest episodes of their favorite series pre-fetched on their mobile phones. More generally, opportunistic or delay-tolerant networks can exploit user mobility to increase an ad hoc network's capacity~\cite{GrossglauserTse02}. However, uncertain delays and probabilistic delivery ratios make such approaches unsuitable for most applications.

Cooperation between the wireless infrastructure and opportunistic networks is a hot topic that has begun to receive attention in the past couple of years. Hui et al. examine how hybrid infrastructure-opportunistic networks can improve delivery ratios over using either paradigm alone. In particular, they show that even infrastructure networks with high access point density can still significantly benefit from the opportunistic capabilities of its users~\cite{hui2009empirical}. Using the wireless infrastructure as a control channel was first suggested by Oliver who exploits the low-cost of SMS to send small messages between participants in an opportunistic mobile network~\cite{oliver2008exploiting}.

Ioannidis et al. push updates of dynamic content from the infrastructure to subscribers that then replicate it epidemically~\cite{ioa09}. The authors assume that the infrastructure has a maximum rate that it must divide among the subscribers. They then calculate the optimal rate allocation for each user in order to maximize the average freshness of content among all subscribers. Han et al. investigate different strategies to find the subset of opportunistic users that will lead to the greatest infection ratio by the end of a message's lifetime. Therefore, pushing the content trough the cellular infrastructure to that optimal subset minimizes the load on the infrastucture~\cite{hui_offloading}. These two papers are close to ours but differ in the following ways. Firstly, they do not have a feedback loop and cannot quickly react to changes in network dynamics or the arrival of new nodes. Secondly, the methods developped in both papers assume preexisting knowledge of pairwise contact probabilities. %

\section{Conclusions}
\label{sec:conclusion}

Push-and-Track is a framework for massively disseminating content with guaranteed delays to mobile users while minimizing the load on the wireless infrastructure. It leverages ad hoc communication opportunities, tracks the content spread through user-sent acknowledgments, and, if necessary, reinjects copies to nodes that have not yet received the content. Tests on the large-scale Bologna vehicular dataset reveal that Push-and-Track manages to reduce the infrastructure load by over 90\% while achieving 100\% delivery. Furthermore, sending small numbers of initial copies lightens the infrastructure load even under tight delay constraints. Finally, pushing content to random nodes works well as it manages to both hit the large connectivity clusters with high probability and spread the pushes uniformly around the city.

Our work will continue in the following directions. Firstly, the feedback loop could be improved, perhaps equipped with a predictive epidemic propagation model. Perhaps the feedback loop could also take into account propagation measurement of previous messages to adjust its strategy for subsequent ones. Secondly, the impact of intermittent infrastructure connectivity must also be explored. Thirdly, any real-life deployement will necessarily be partial and progressive. How does Push-and-Track fare when only a fraction of all users participate? Finally, this paper dealt with the case where all users were interested in the same content. However, the Push-and-Track framework is flexible and can be extended to a more realistic setting in which overlapping subsets of users concurrently request different content.

\section*{Acknowledgments}
{\small
We would like to especially thank the iTETRIS partners that have made available and built the vehicular dataset. For this, we especially thank Fabio Cartolano, Carlo Michelacci, and Antonio Pio Morra from the Municipality of Bologna, as well as Daniel Krajzewicz from the German Aerospace Center. We also thank Javier Gozalvez, Ramon Bauza, Cl\'emence Magnien, and Matthieu Latapy for their comments.
This work has been partly funded by the European project iTETRIS (No. FP7 224644) and the French ANR CROWD project under contract ANR-08-VERS-006.}

\bibliographystyle{IEEEtran}
\bibliography{xbib_itetris}

\end{document}